\documentclass{article}
\usepackage[utf8]{inputenc}

\usepackage{graphicx}
\usepackage{algorithm}
\usepackage{algorithmicx}
\usepackage{algpseudocode}
\usepackage{authblk}

\algnewcommand{\algorithmicand}{\textbf{ and }}
\algnewcommand{\algorithmicor}{\textbf{ or }}
\algnewcommand{\OR}{\algorithmicor}
\algnewcommand{\AND}{\algorithmicand}
\algnewcommand{\var}{\texttt}

\title{The social graph based on real data\footnote{Preprint of paper presented at ICCS'23, https://doi.org/10.1007/978-3-031-36027-5\_1}}
\author{Tomasz M. Gwizdałła\thanks{tomasz.gwizdalla@uni.lodz.pl} }
\author{Aleksandra Piecuch }
\affil{Faculty of Physics and Applied Informatics, \\University of Lodz, Lodz, Poland}
\date{}
\begin{document}

\maketitle              

\begin{abstract}
In this paper, we propose a model enabling the creation of a social graph corresponding to real society. The procedure uses data describing the real social relations in the community, like marital status or number of kids. Results show the power-law behavior of the distribution of links and, typical for small worlds, the independence of the clustering coefficient on the size of the graph.\end{abstract}


\section{Introduction and Model}

The computational study of different social processes requires knowledge of the structure of the community studied. 
Among the seminal papers devoted to study of graph representations of communities we have especially to mention the Pastor-Satorras paper \cite{Pastor_15} where different properties of graphs are studied in the context of epidemic. Now, a lot of problems can be studied with the use of social graphs. We can mention here the processes which can be generally called social contagion \cite{Arruda_23}. Particularly, such topics like disease spreading \cite{Karaivanov_20,TMG_20_CMPB,Correia_23}, transfer of information \cite{Kumar_21}, opinion formation \cite{Kozitsin_22,Morales_23} or innovation emergence \cite{Sziklai_23} are studied with the use of particular graphs. It is important to emphasize that speaking about social network in the context of studied problem we have take into account the real community which structure does not necessarily correspond to the structures obtained by studying, recently very popular, online networks. The technique which have recently become popular for such networks is the study of homophily effect \cite{Talaga_20,ZhaoTong_14}. 
In our paper, we will present a new approach to creating a society graph, based on the real statistics describing it. The special attention is devoted to study the real social links like families, acquaintances. Due to ease of access, we use the data provided for Polish society, mainly our city - Lodz.

The population is typically presented as an undirected, connected graph, where nodes correspond to individuals and edges to relations between them. Both features mentioned above are essential for the understanding of the model. When assuming the graph's undirected character, we consider the mutual type of interaction. It means that compared to well-known social networks, we consider rather a connection (like the most popular way of using linkedin) than observation (like twitter). When assuming the connected graph, we ensure that, in society, we do not have individuals that are completely excluded. We will come back to this issue when discussing the levels of connections.

We propose to use initially four levels of connection between individuals/nodes
\begin{itemize}
\item {\bf I.} Household members and inmates. In this class, we collect the people who have permanent contact - members of families, cohabitants, or people who rent apartments together.
\item {\bf II.} In this class, several types of connections are considered. The crucial property of people in these groups is that they are strongly connected but outside households. We include small groups created in the school or work environment and further family contacts here. These people meet themselves every day or almost every day, and small social distances characterize their meetings.
\item {\bf III.} Acquaitances. In this class, we consider loosely related individuals but having the possibility of frequent contacts - kids in schools or employees in companies.
\item {\bf IV.} Accidental but possible contacts - here we sample mainly the people living in the common areas - villages, city districts. We can meet them accidentally on the street, in the shop, but these meetings are strongly random.
\end{itemize}
In the paper, we focus on the first two categories. However, the creation of a third class will also be described because the second one is based on it. We assume that at least one individual is always considered as a family member (1st class) or caregiver (2nd class).

To ensure the correctness of the data, we use only the ones which follow some basic criteria.
\begin{itemize}
\item The source of data can be considered reliable. This criterion, indeed, led to the strong restriction that all data come from the polish GUS (Statistics Poland - https://stat.gov.pl)
\item The studies were conducted in a similar period. It is tough to find all data sampled simultaneously during the same study because they cover different aspects of social or economic statistics.
\item The studies were conducted on a similar statistical population. Sometimes, the easily accessible data concern different communities, e.g., big cities and village populations. In such a situation, we do rather consider the global data for the whole country (like the employment vs. size of company statistics).
\end{itemize}
In the paper, we model the closed community corresponding rather to a big city than to scattered areas of villages and small towns. Such a choice is made for several reasons. The regulations for such environments are more regular, without many exceptions, like e.g., the size of schools. The modeling of the distribution of individuals, for example, between districts, is also easier than creating a large number of small communities, where additional conditions should be taken into account (like e.g., the general possibilities of connecting between such communities). The data presented below differ significantly for various communities. In the calculations, we mainly use the data for our city - Lodz ({\L}\'od\'z), or for the whole of Poland.

The successive steps of the proposed approach is as follows:
\begin{itemize}
\item We start by selecting the correct number of population members according to the age pyramid. Here we use the data for Lodz, and the borders of intervals for which data are known are given by $\{2, 6, 12, 15, 18, 24, 34, 44, 54, 64, 75, 95\}$. For a relatively small percentage of people in the last interval (above 95), we merge the last two intervals.
\item The marital status of men is sampled. We use the data concerning the five possible states: bachelor, married, cohabitant, divorced, and a widower for different age intervals (different from the age pyramid). We start from the men subset because the total number of men is smaller than the number of women.
\item The marriages and cohabitation relations are created as the first level connection. We use the distribution of differences between men's and women's ages in relations and sample the partners from the women's subset. For the remaining women, we attach the status in the same manner as for men. The correctness of the final marital distribution is the test for the correctness of the whole procedure, but we do not present the results confirming this effect here.
\item The kids are attached as the first-level connection. Having the distribution of kids typical for a particular age and the type of family (pair, single men, single woman), we attach the number of kids of an age different from the age of the woman (or man) by a value from interval [18,40].
\item The kids and youth are attached to schools as the 3rd-level connections. We know the average size of a school of a particular type, so we can, taking into account the number of individuals in the corresponding age, determine the number of schools. Then we create cliques, attaching every individual under 18 (in accordance with Polish law) to the one of the established schools.
\item People working in the same companies are connected as the 3rd-level connections. The basis for assigning people to jobs is the distribution of a number of employees in one of the four groups specified by polish statistics (micro, small, average, big - sometimes also the group of large companies is described, but we omit it). After assigning the type of company to every individual of the right age (before the retirement age, different for men and women), we create cliques of size sampled uniformly from the interval characteristic for particular group.
\item The selector of narrower groups as the 2nd-level connections. We choose the smaller cliques among the 3rd-level connections. We can choose them arbitrarily, but for the presented calculations, we use the Poisson distribution with an average equal to $3$ as the size of these cliques.
\item Finalization. After the steps described above, there can still be some unconnected nodes. It mainly concerns retired people, who are not in pairs at the first level, are not in a household with children and do not work professionally. For these people, we attach the person, younger by 20-40 years, who can be considered as a separately living child or a caregiver.
\end{itemize}
The procedure described above allows the creation of the connected graph when the first and second-level connections are considered. Intentionally, we do not describe the procedure of creating connections belonging to the 4th-level because it is not analyzed in the paper.

\begin{figure}[!h!]
\includegraphics[width=0.9\textwidth]{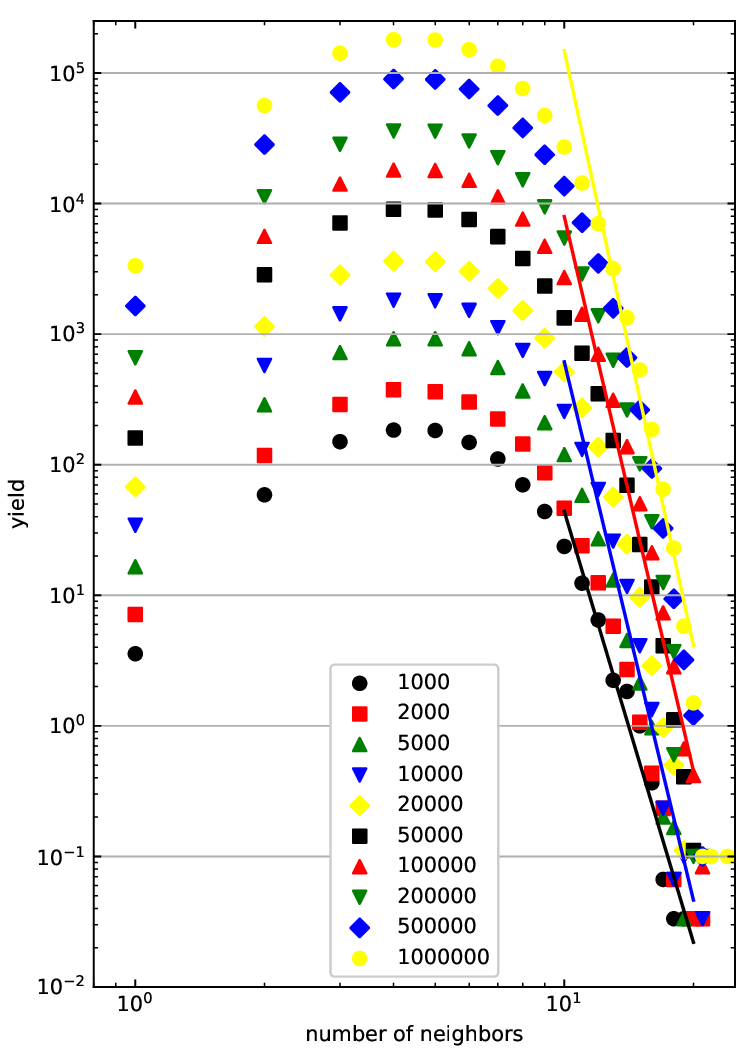}
\caption{The comparison of links distribution histograms for different graph sizes. The solid lines, created for selected set of graph sizes ($10^3$, $10^4$,$10^5$,$10^6$) correspond the power-law dependencies and the colors correspond to the color of marker for particular size of community. }
\label{f3_hist}
\end{figure}

\section{Results and Conclusions}

We concentrate our attention on four parameters that are typically used to describe the properties of a graph as the form representing the community. They are the number of neighbors/links distribution, radius and diameter, and clustering coefficient. The results are presented for graph sizes from 1000 to 1000000, distributed uniformly in the log scale. The calculations for every number are performed several times, and the presented values are averaged over independent runs. The number of repetitions varies from 30 for the smaller sizes to 5 for greater ones. Due to large computational complexity we do not present results for radius and diameter for largest graphs. The spread of the averaged values is presented on plots by the bar corresponding to the standard deviation value.

The distribution of the number of links is shown in Fig.\ref{f3_hist}. The main observation is that our procedure does not produce hubs. It is clear since in the presented scheme, we try to create, as the first and second level connections, cliques with the size distribution sampled from the Poisson distribution with a relatively small average - $3$. The crucial property is, in our opinion, that for our approach, we observe the power-law scaling for the descending part of the plot. The value of the exponent for the presented cases shows interesting behavior. They are visibly larger than the typical for BA value $3$; they start from about $8$ for 1000, seem to exceed $11$, and start to decrease to about $10$ for 100000. This effect needs, however, more detailed analysis.

\begin{figure}[!h!]
\begin{center}
\includegraphics[width=0.8\textwidth]{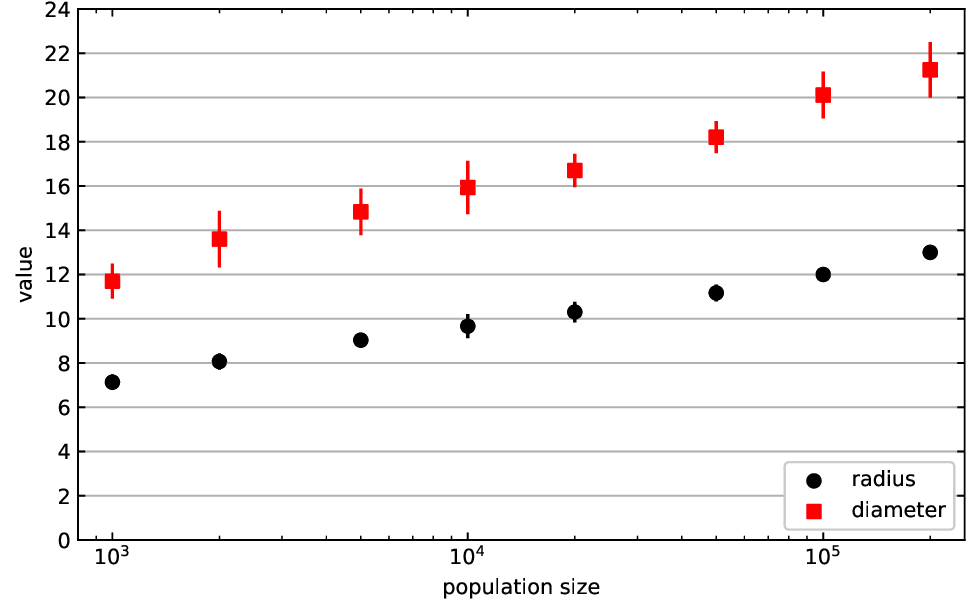}
\end{center}
\caption{The dependence of radii and diameters on the size of community for graphs created according to the presented procedure. }
\label{f1_radius}
\end{figure}

The dependence of radii and diameters of graphs on the graph size, shown in Fig.\ref{f1_radius}, presents the typical, generally logarithmic character. We can observe that the dispersion of diameters is larger than radii and that the values are larger than the predicted for scale-free Barabasi-Albert network \cite{Bollobas_04}.

\begin{figure}[!h!]
\centering
\includegraphics[width=0.8\textwidth]{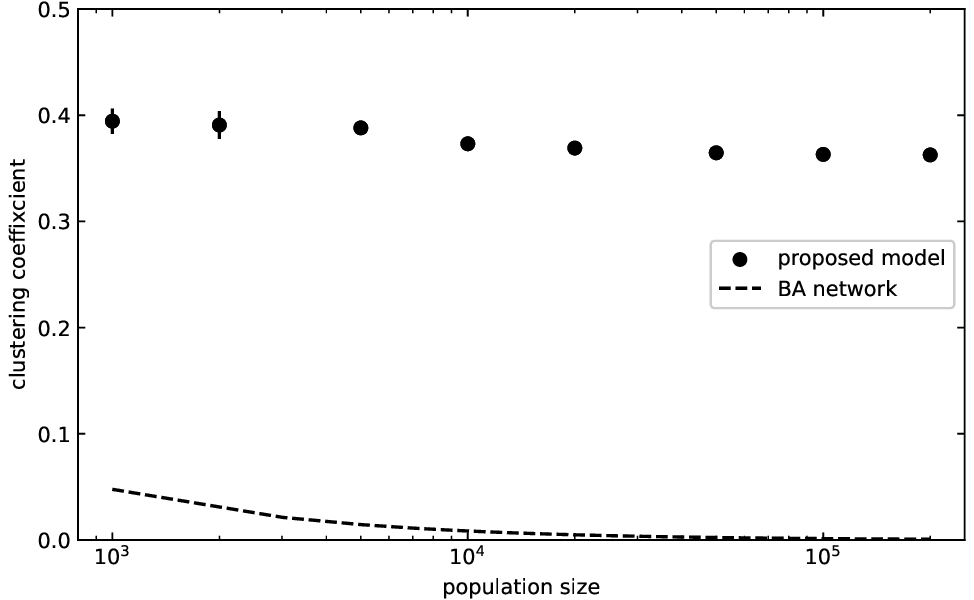}
\caption{The dependence of clustering coefficient on the size of community for graphs created according to the presented procedure. The values were obtained in the same wat as those from Fig.\ref{f1_radius}.}
\label{f2_cc}
\end{figure}

A significant effect can be observed in the plot of clustering coefficient \ref{f2_cc}. For comparison, on the plot, we also present the curve presented for scale-free Barabasi-Albert network \cite{Klemm_02,Bollobas_03}. ??? It is essential that for our network we obtain, the expected for small-worlds constancy of the coefficient. Only very small descend is obseved, but its magnitude is of the order of a few percent.

The presented results show that the proposed model leads to interesting and promising effects. The crucial ones are the power-law distribution of links and the constant value of the clustering coefficient. We can easily indicate the the directions in which the model should be developed. We have to consider the appropriate way to add hubs; we have to think about the change of distribution of close acquaintances (2nd level), and we also think about another organization of connection levels. However, the changes mentioned above can be, in our opinion, considered as small corrections and will not substantially affect the properties shown in the paper.

\bibliographystyle{splncs04}
\bibliography{biblio}

\end{document}